# Significant $CO_2$ photoreduction on a high-entropy oxynitride

Saeid Akrami[1], Parisa Edalati[1], Yu Shundo[2,3], Motonori Watanabe[2], Tatsumi Ishihara[2,3,4], Masayoshi Fuji[1,5] and Kaveh Edalati[2,3,*]

[1] Department of Life Science and Applied Chemistry, Nagoya Institute of Technology, Tajimi 507-0071, Japan

[2] WPI, International Institute for Carbon-Neutral Energy Research (WPI-I2CNER), Kyushu University, Fukuoka 819-0395, Japan

[3] Mitsui Chemicals, Inc. - Carbon Neutral Research Center (MCI-CNRC), Kyushu University, Fukuoka 819-0395, Japan

[4] Department of Applied Chemistry, Faculty of Engineering, Kyushu University, Fukuoka 819-0395, Japan

[5] Advanced Ceramics Research Center, Nagoya Institute of Technology, Tajimi 507-0071, Japan

**Abstract**

$CO_2$ photoreduction on photocatalysts is a nature-friendly solution to decrease the $CO_2$ amount, but the method still has low efficiency because of difficult separation and easy recombination of charge carriers in available catalysts. In this study, a high-entropy oxynitride was introduced as an active photocatalyst for photoreduction. The material had a chemical composition of $TiZrNbHfTaO_6N_3$ and was produced by a high-pressure torsion method followed by oxidation and nitriding. It showed higher photocatalytic $CO_2$ to CO conversion compared to corresponding high-entropy oxide, benchmark photocatalyst P25 $TiO_2$, and almost all catalysts introduced in the literature. The high activity of this oxynitride, which also showed good chemical stability, was attributed to the large absorbance of light and easy separation of electrons and holes, the low recombination of charge carriers, and the high $CO_2$ adsorption on the surface. These findings introduce high-entropy oxynitrides as promising photocatalysts for $CO_2$ photoreduction.

*Corresponding author:
    Kaveh Edalati  (E-mail: kaveh.edalati@kyudai.jp; Tel: +81-92-802-6744)

## 1. Introduction

Global warming by significant $CO_2$ emission is a universal concern in recent years which has forced scientists and politicians to find a remedy to convert this harmful gas into useful substances [1,2]. $CO_2$ photoreduction as an artificial photosynthesis method is a clean strategy that can transform $CO_2$ into reactive or valuable components like CO and $CH_4$ [1,2]. $CO_2$ photoreduction takes place on the surface of a photocatalyst (usually a semiconductor) under irradiation by solar light. Electrons in the valence band of the photocatalyst absorb the light and transfer into the conduction band and misplace the holes within the valence band [1,2]. Exited holes and electrons take part in reactions for oxidation and reduction, respectively, and produce CO, $CH_4$ and other hydrocarbons [1,2]. The challenge in this field is to discover an appropriate photocatalyst with high light absorbance, low bandgap, appropriate band positions, a low recombination rate of electrons and holes, high $CO_2$ adsorbance and high chemical stability [3,4].

Oxide photocatalysts such as $TiO_2$ [5,6] are the most common photocatalysts with high stability for $CO_2$ photoreduction application, but they have large bandgaps such as 3.1 eV for $TiO_2$ [5,6] In contrast, there are some reports on photocatalytic activity of nitrides such as TaN [7] and $C_3N_4$ [8] for $CO_2$ conversion which have lower bandgaps compared to oxides, but nitrides are not chemically so stable [7,8]. To solve the problem of oxide and nitride photocatalysts in terms of large bandgap and low stability, respectively, oxynitrides were recommended as low bandgap and highly stable catalysts for photocatalysis [9]. Oxynitrides have been widely used for photocatalytic water splitting; however, only limited oxynitrides such as $\alpha$-$Fe_2O_3$/$LaTiO_2N$ [10] and TaON [11,12] were used for photocatalytic $CO_2$ conversion. Significant electron-hole recombination, sluggish kinetics, the low tendency for $CO_2$ adsorption and relatively low stability in the co-presence of $CO_2$ and water are some reasons for limited application of metal oxynitrides for photocatalytic $CO_2$ conversion [10,13]. Therefore, introducing a strategy to solve all or some of these problems is a key issue in using the benefits of oxynitrides for $CO_2$ conversion. Simultaneous addition of several principal elements and production of high-entropy oxynitride ceramics can be an effective strategy, although there have been few attempts in this regard.

High-entropy ceramics containing at least five cations and having a configurational entropy of larger than $1.5R$, where $R$ is the molar gas constant, are new remarkable candidates for various applications due to their superior stability, large lattice defects/strain and heterogenous valence electron distribution [14,15]. Among various kinds of high-entropy ceramics, high-entropy oxides (HEOs) have become quite popular due to their feasibility for various applications. Catalysis is a new application field for highly stable HEOs which has been expanded in recent years as a top issue [16]. These oxides have been used as electrocatalyst for oxygen evolution [17-19], catalyst for CO oxidation [20-22], catalyst in lithium-sulfur batteries [23], electrocatalyst for $CO_2$ conversion [24], electrocatalyst in electrochemical capacitors [25], catalyst for combustion reactions [26], and photocatalyst for redox reactions [27,28]. High entropy nitrides (HENs) are another popular type of high-entropy ceramics and have been used as coatings [29], supercapacitors [30] and solar selective absorbers [31]. A combination of the perception of metal oxynitrides as low bandgap photocatalysts and high-entropy ceramics as highly stable materials can be a new strategy to expand the application of oxynitrides for photocatalytic $CO_2$ conversion. Although high-entropy oxynitrides (HEONs) were successfully synthesized in a few studies [32,33], there are no reports on the photocatalytic performance of HEONs for $CO_2$ photoreduction.

In this study, a two-phase $TiZrNbHfTaO_6N_3$ was synthesized as the first HEON for photocatalytic $CO_2$ conversion by high-pressure torsion mechanical alloying [34] and subsequent oxidation and nitriding. The HEON showed better light absorbance, lower charge carrier recombination rate, higher $CO_2$ adsorbance and larger photocatalytic $CO_2$ conversion compared to relevant HEO ($TiZrNbHfTaO_{11}$) and benchmark photocatalyst P25 $TiO_2$. Moreover, the activity of the HEON was higher than almost all photocatalysts developed in the literature for $CO_2$ photoreduction. These findings open a path to develop new high-entropy photocatalysts with significant efficiency for $CO_2$ photoreduction.

## 2. Experimental procedures

Despite various synthesis methods reported in the literature to produce HEOs [14-28] and HENs [14,15,29-31], the HEON was fabricated using a three-step synthesis method for this study [33]: (i) severe plastic deformation through the high-pressure torsion (HPT) method for alloying pure elemental powders [34,35], (ii) oxidation at elevated temperature and (iii) nitriding at elevated temperature. First, titanium (99.9%), zirconium (95.0%), niobium (99.9%), hafnium (99.5%) and tantalum (99.9%) powders with the same molar fraction of 0.2 were dispersed in acetone, mixed using ultrasonic and dried in air. About 700 mg of powder mixture was compacted into a 10 mm diameter disc under a pressure of 0.4 GPa and further proceeded by HPT under 6 GPa at room temperature using a rotation rate of one turn per minute for 100 turns to achieve a single-phase (body-centered cubic, BCC) alloy. Second, the HPT-processed high-entropy alloy was crushed in a mortar and inserted into a furnace for 24 h under a hot (1373 K) air atmosphere to generate the HEO, $TiZrNbHfTaO_{11}$ with dual monoclinic (40 wt%) and orthorhombic (60 wt%) structures. Third, the HEO was processed by nitriding in ammonia at 1373 K for 7 h using a heating rate of 20 $Kmin^{-1}$ with an $NH_3$ flow of 150 $mLmin^{-1}$ to generate a two-phase (40 wt% monoclinic + 60 wt% face-centered cubic, FCC) HEON, $TiZrNbHfTaO_6N_3$. The fabricated HEON was characterized by different methods, as follows.

The crystallographic features were analyzed by X-ray diffraction (XRD) using a Cu Kα source having 0.1542 nm wavelength. Phase fractions and lattice parameters were measured by the Rietveld analysis in the PDXL software.

The composition was examined by dispersing the sample on a carbon tape and conducting energy-dispersive X-ray spectroscopy (EDS) in a scanning electron microscope (SEM) under 15 keV.

The microstructure was examined by dispersing the crushed sample on carbon grids and employing a transmission electron microscope (TEM) under 200 keV by taking high-resolution (HR) images and analyzing them by fast-Fourier transform (FFT). Moreover, the distribution of elements was examined by a scanning-transmission electron microscope (STEM) under 200 kV by taking high-angle annular dark-field (HAADF) micrographs and conducting EDS analysis.

X-ray photoelectron spectroscopy (XPS) was performed to determine the top of the valence band and the electronic state of each element using a Mg Kα source.

The absorbance of light, and band structure including the level of bandgap were evaluated by UV-vis diffuse reflectance spectroscopy (followed by Kubelka-Munk calculation) and X-ray/UV photoelectron spectroscopy (XPS and UPS). The valence band top was determined by the UPS and XPS analyses and the conduction band bottom was determined by subtracting the bandgap value from the valence band top.

The electron-hole recombination was evaluated by photoluminescence (PL) spectroscopy using a UV laser (325 nm wavelength).

Photocurrent measurement on thin films of the samples was performed using the full arc of a Xe lamp in a 1 M $Na_2SO_4$ electrolyte. The experiments were conducted in the potentiostatic amperometry mode during the time (180 s light ON and 180 s light OFF) using an electrochemical analyzer. The counter and reference electrodes were Pt wire and Ag/AgCl, respectively, and the external potential was 0.7 V vs. Ag/AgCl. To prepare the thin films, 5 mg of each sample was crushed in 0.2 mL ethanol, spread on the FTO glass (fluorine-doped tin oxide with 2.25 mm thickness and $15 \times 25$ mm$^2$ surface area), and annealed at 473K for 2 h.

Diffuse reflectance infrared Fourier transform (DRIFT) spectrometry was performed to understand the adsorbance mode of $CO_2$ on the surface of each photocatalyst. First, 50 mg of each sample was treated at 773 K for 1 h in an argon atmosphere. Then, argon was replaced by 100% $CO_2$ gas at 773 K and the samples were kept under this condition for 30 min. The samples were then cooled down to room temperature and the $CO_2$ gas was replaced with argon. After keeping the samples in argon for 30 min, the DRIFT spectroscopy was conducted.

$CO_2$ photoreduction was examined in a cylindrical-shaped quartz photoreactor with 858 mL inner volume and specifications described in detail earlier [28]. Light source was placed in a space inside the photoreactor and $CO_2$ flow entered the reactor from a gas cylinder by a hole on the top of the reactor. Outlet gas from the reactor partly entered a gas chromatograph for the gas analysis and mainly flew to a vent. For the photoreduction experiments, 100 mg of HEON photocatalyst were dispersed in a 500 mL solution of 1 M NaHCO$_3$ and pure water. $CO_2$ gas was injected into the mixture (3 mLmin$^{-1}$) and the mixture was continuously stirred by a magnetic stirrer. It should be noted that the temperature was kept constant at 288 K utilizing a water chiller. To be sure about the nonappearance of reaction products without irradiating the mixture, the experiments were first performed for 2 h in dark conditions. Then the photocatalytic experiment was performed under irradiation of a 400 W high-pressure mercury lamp (HL400BH-8 of Sen Lights Corporation) with 0.5 Wcm$^{-2}$ light intensity without any filtration. The gas of the photoreactor was analyzed using gas chromatography (GC-8A of Shimadzu). The generation of $CH_4$ and CO was analyzed using a methanizer and flame-ionization detector. The generation of oxygen and hydrogen was analyzed using a thermal conductivity detector. A blank test was conducted without catalyst addition under light irradiation and $CO_2$ injection to confirm that no CO was produced from other sources in the experimental system. Another blank test was conducted with catalyst addition under light irradiation and argon injection to confirm that CO was not produced without $CO_2$ injection.

## 3. Results

Fig. 1a illustrates the XRD profile of HEON. The HEON has two cubic (Fm3m space group, $a=b=c=0.459$ nm; $\alpha=\beta=\gamma=90°$) and monoclinic (P21/c space group, $a=0.512$, $b=0.517$, $c=0.530$ nm; $\alpha=\gamma=90°$, $\beta=99.2°$) phases with 60 and 40 wt% fractions, respectively. Fig. 1b illustrates the EDS profile of the HEON. The EDS analysis suggests a general composition of TiZrHfNbTaO$_6$N$_3$ for the synthesized HEON. The presence of two phases should be due to the thermodynamics of the Ti-Zr-Hf-Nb-Ta-O-N system at the synthesis temperature. The existence of two phases can be beneficial for charge separation in photocatalysis because the phase boundaries can act as heterojunctions for charge carrier migrations [36,37]. Although first-principle electronic structure calculations are required to clarify the migration direction of charge carriers in this HEON, it is expected that photoexcited electrons in the conduction band of one phase with a higher energy level move to the conduction band of the other phase and exited holes transfer from the valence band of one phase with the lower energy level to the valence band of another phase [36,37].

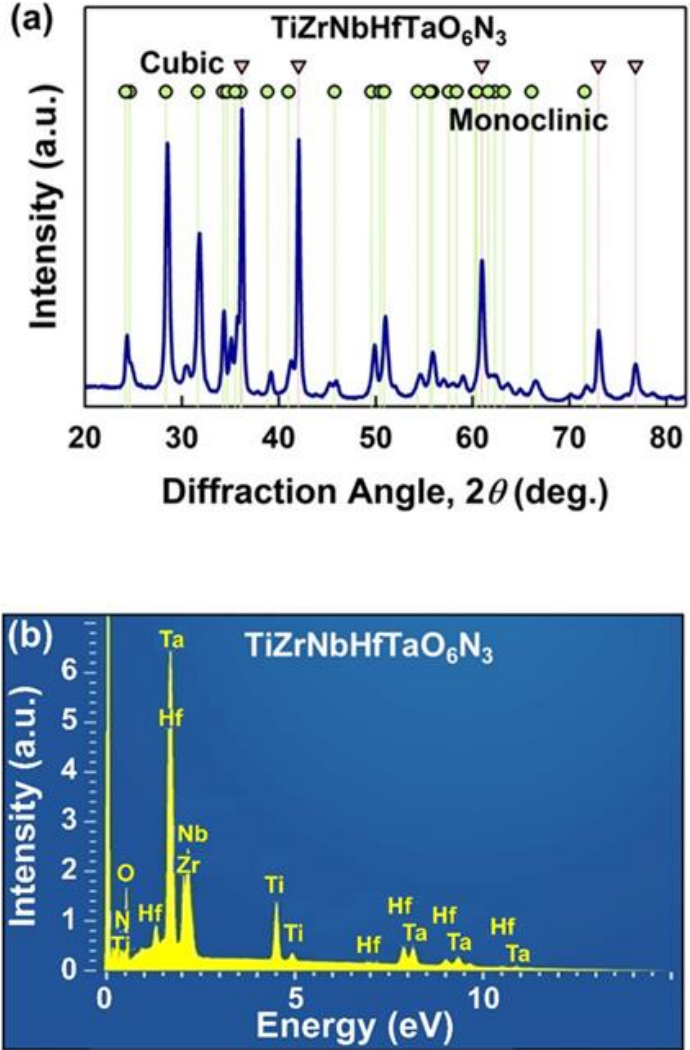

Fig. 1. Formation of high-entropy oxynitride with cubic and monoclinic phases and chemical composition of TiZrNbHfTaO$_6$N$_3$. a) XRD profile and b) EDS spectrum of high-entropy oxynitride.

The microstructure of the HEON is shown in Fig. 2 using different methods. Fig. 2a illustrates a micrograph taken by SEM, which indicates that the HEON contains large powders with an average size of 20 µm. Fig. 2b shows a HR image taken by TEM confirming the existence of nanocrystals of cubic and monoclinic phases which agrees with the XRD analysis. It also indicates the existence of a large fraction of interphases that can act as heterojunctions [1]. Fig. 2c illustrates a HAADF micrograph with relevant EDS mappings taken by STEM, showing a reasonably homogenous distribution of elements at the nanometer scale. Slight differences in the distribution of metallic elements, oxygen and nitrogen should be mainly due to the presence of two phases. Here, it should be noted that XPS analyses, shown in Supporting Information Fig. S1, confirm that the main states of elements are Ti$^{4+}$, Zr$^{4+}$, Hf$^{4+}$, Nb$^{5+}$, Ta$^{5+}$, O$^{2-}$ and N$^{3-}$.

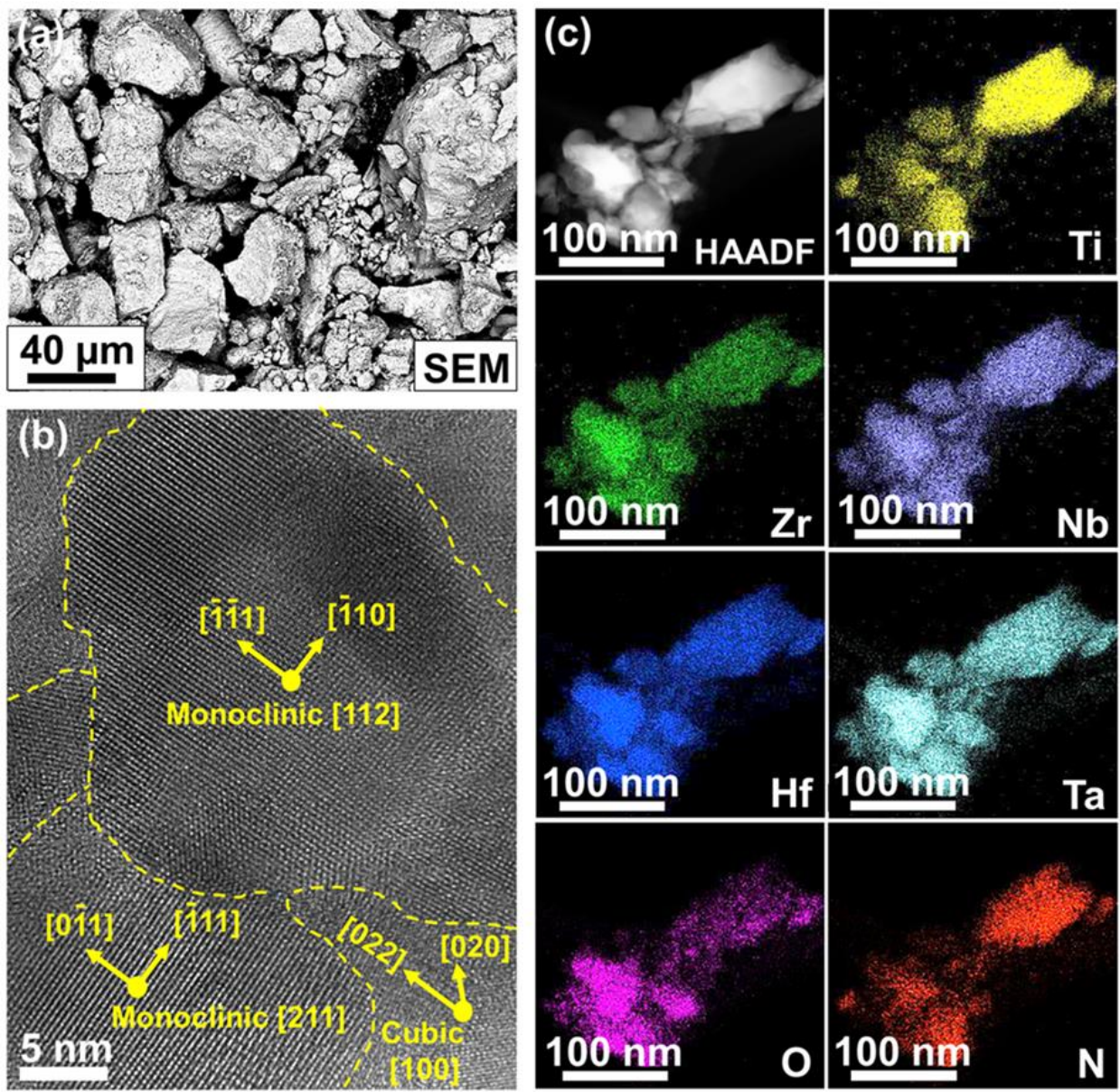

 Formation of nanocrystalline monoclinic and cubic phases with uniform elemental distribution in high-entropy oxynitride powder. a) SEM micrograph, b) HR micrograph and c) STEM-HAADF micrograph and relevant EDS mappings for high-entropy oxynitride.

Fig. 3a shows the light absorbance of the HEON in comparison with the relevant HEO as well as the P25 TiO$_2$ photocatalyst. The HEON exhibits significant light absorbance compared with the HEO and P25 TiO$_2$. According to the Kubelka-Munk calculation, the bandgap for the HEON is 1.6 eV which is extremely narrower compared with the bandgap of the HEO (3.0 eV) and P25 (3.1 eV) [28]. Fig. 3b shows the electronic band structure of the three mentioned materials including the appearance of three samples. A color change from white and orange for P25 TiO$_2$ and the HEO occurs to dark brown for the HEON, confirming the high light absorbance of the HEON in good agreement with the UV-vis absorbance data [38]. The electronic band structures were determined by considering the bandgaps calculated using the Kubelka-Munk theory, the top of the valence band was measured by XPS spectroscopy, and the bottom of the conduction band was calculated by subtracting the bandgap from the top of the conduction band. The bandgap for the P25 $TiO_2$, the HEO and the HEON are 3.0, 3.0 and 1.6 eV, respectively; the values for the top of the valence band are 2.2, 1.8 and 1.3 eV vs. NHE for P25 $TiO_2$, the HEO and the HEON, respectively; and the values for the bottom of the conduction band are -0.8, -1.2 and -0.3 eV vs. NHE for P25 $TiO_2$, the HEO and the HEON, respectively. As shown in Fig. 3b, the band structure of the HEON indicates its low bandgap with appropriate positions of the valence band top and the conduction band bottom for various $CO_2$ conversion reactions [39]. The low bandgap of this HEON can lead to easy separation of electrons and holes during photocatalysis.

Fig. 3c shows the photoluminescence spectra of the three materials to examine the electron-hole recombination. P25 $TiO_2$ and the HEO have almost the same photoluminescence intensity while the HEON shows the lowest photoluminescence. The absence of an intensive photoluminescence peak for the HEON confirms the significant suppression of electron-hole recombination which is a principal requirement for the enhancement of photocatalytic reactions [2,3]. These results show that the main problem of metal oxynitrides in terms of high electron-hole recombination [11,13] can be solved by the strategy used in this study through the concept of high-entropy ceramics.

Fig. 3d shows the photocurrent measurement for the three samples. Due to the different particle sizes of these three samples and their dissimilarities in making binding to the FTO glass, their current density cannot be compared quantitatively; however, the shape of their photocurrent curves can clarify their different behaviors. For P25 $TiO_2$ there is a spike peak at the beginning of irradiation, but the current density decreases rapidly to a steady state, suggesting that electron-hole separation is followed by fast recombination. For the HEO the photocurrent curve under irradiation is almost a straight horizontal line which shows a better electron-hole separation of the HEO compared to P25 $TiO_2$. For the HEON, the current density increases by irradiation and reaches a steady state, suggesting that the ratio of electron-hole recombination to separation is the lowest for the HEON. After stopping the irradiation, the HEON still shows some reduced photocurrent due to the remained excited charge carriers, while P25 $TiO_2$ exhibits almost no photocurrent under the dark condition. The successful photocurrent generation on this HEON with an appropriate ratio of electron-hole separation to recombination suggests the potential of this material to act as a catalyst for $CO_2$ photoreduction [28,38].

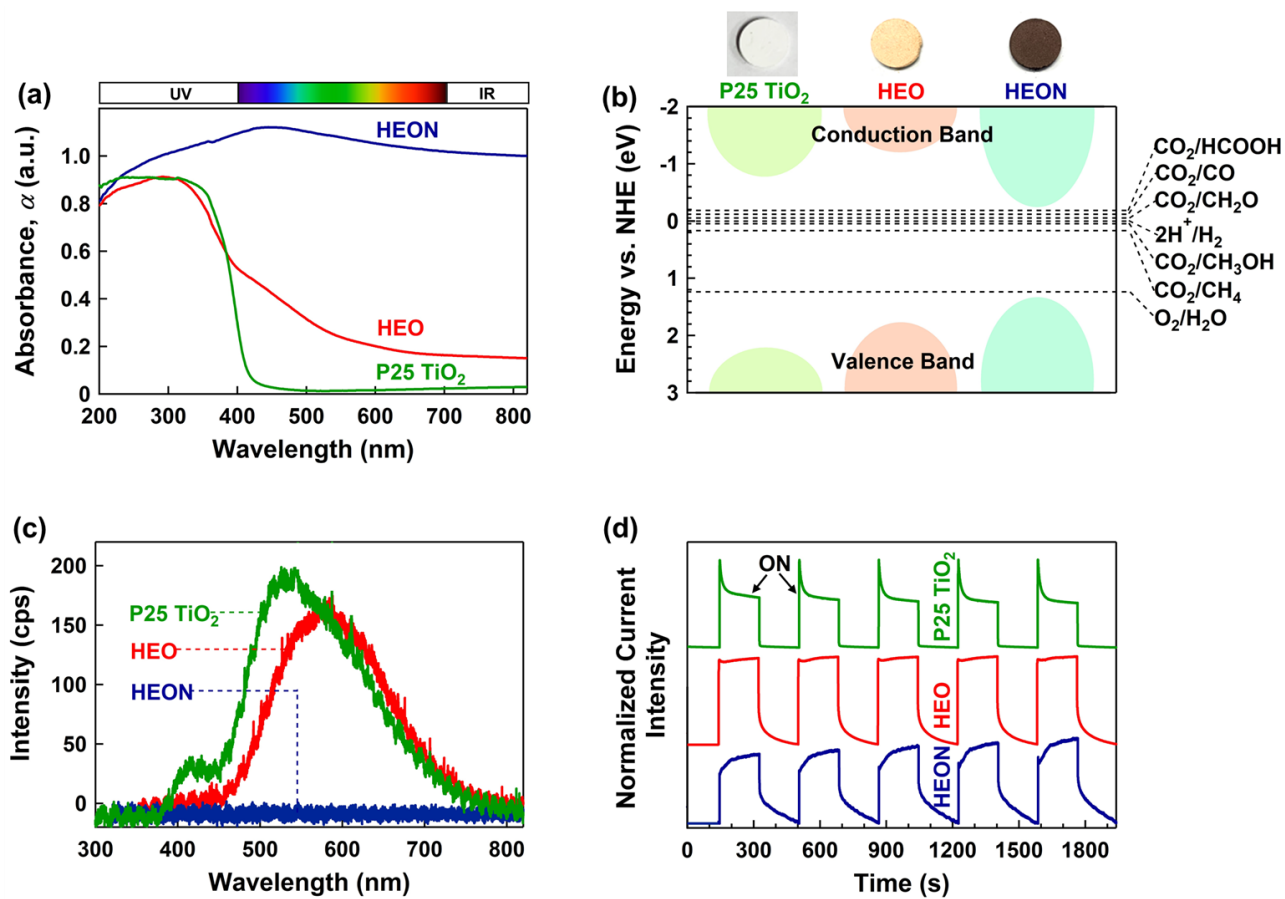

Fig. 3. High light absorbance, appropriate electronic band structure and low electron-hole recombination in high-entropy oxynitride. a) UV-vis spectra, b) electronic band structures including chemical potential for $CO_2$ conversion reactions and sample color, c) steady-state photoluminescence spectra and d) photocurrent generation of high-entropy oxynitride (HEON) in comparison with corresponding high-entropy oxide (HEO) and P25 $TiO_2$.

Fig. 4a compares the CO production rate of the HEON with the relevant HEO and P25 $TiO_2$ benchmark photocatalyst. Photoreduction of $CO_2$ to CO on the HEON is considerably better than the HEO and P25. The CO production rate for the HEON reaches 14.3 µmolh$^{-1}$g$^{-1}$ after 1 h; and then decreases to a constant level after 3 h. The deviations in the reaction rate in the first two hours should be due to the time needed to reach an equilibrium condition in the measurement system. The average CO production rate for this HEON is $11.6 \pm 1.5$ µmolh$^{-1}$g$^{-1}$ after 5 h, while the HEO and P25 have lower photocatalytic CO production rates of $4.6 \pm 0.3$ µmolh$^{-1}$g$^{-1}$. Fig. 4b shows the photocatalytic activity of the HEON for $H_2$ production and compares it with the HEO and P25 $TiO_2$. This figure indicates that the efficiency of the HEON is much better than the HEO and P25 for photocatalytic $H_2$ production. The average $H_2$ production rate for the HEON is $5.1 \pm 0.5$ µmolh$^{-1}$g$^{-1}$, while the rate for the HEO and P25 $TiO_2$ is $1.3 \pm 0.1$ and $1.5 \pm 0.1$ µmolh$^{-1}$g$^{-1}$, respectively. To confirm the high activity and stability of this HEON for $CO_2$ photoreduction, a long-term photocatalytic experiment for 20 h was conducted on the sample after storage in air for 7 months. As shown with dashed-line curves in Fig. 4a and 4b, the material still shows high activity with a constant CO and $H_2$ production rate, although the reaction rates are slightly lower than in the first experiment.

Here, three issues regarding the photocatalytic tests should be mentioned. First, no CO was detected in three blank tests: (i) with catalyst addition and $CO_2$ injection under dark conditions, (ii) without catalyst addition under light irradiation and $CO_2$ injection, and (iii) with catalyst addition under light irradiation and argon injection. Second, despite the high light absorbance of HEON in the visible light and near-infrared region, the material did not show any photocatalytic activity in these regions within the detection limits of gas chromatographs. Third, despite the higher feasibility of $CH_4$ production compared to CO generation in terms of thermodynamics, no $CH_4$ was detected for these materials which can be explained by the kinetics of reactions. Due to the requirement of $CH_4$ production to more electrons and protons, its production is not kinetically more feasible than CO production [40,41]. On the other hand, once CO is produced, it does not tend to be adsorbed on active sites and thus the reaction terminates with the CO production [42].

Fig. 4c illustrates the XRD profiles of HEON before and after photocatalysis. The profiles indicate that the crystal structures do not change after photoreduction, suggesting that the HEON remains stable after photocatalytic $CO_2$ conversion. The high stability of $TiZrHfNbTaO_6N_3$ is partly because of the entropy-stabilization concept which leads to low Gibbs free energy in the presence of a large number of elements [14,15]. This high stability is an important issue that has led to the utilization of high-entropy ceramics for various applications with superior performance [14-33].

Fig. 4d shows the DRIFT spectra for the three samples to investigate the adsorbance mode of $CO_2$ on the surface of each photocatalyst. There is a peak at 665 cm$^{-1}$ which corresponds to $CO_3^{2-}$ [43] and another one at 2340-2360 cm$^{-1}$ which is relevant to $CO_2$ gas in the beamline of spectrometer or to physically adsorbed $CO_2$ on photocatalysts [44]. The intensity of both peaks is the maximum for the HEON, but it is hard to discuss about physically adsorbed $CO_2$ using the peak at 2340-2360 cm$^{-1}$ due to the possible differences in the $CO_2$ gas concentration in the beamline. The peak at 665 cm$^{-1}$ for P25 $TiO_2$ is so weak, but its intensity is the highest for the HEON, suggesting $CO_2$ can bond to the surface as carbonate. Since $CO_2$ is a Lewis acid, the basic active sites have a significant role in the adsorption and activation of this molecule [45]. P25 $TiO_2$ is considered a weak acid and chemisorption of $CO_2$ in the form of carbonate is weak on the surface of this material. The high intensity of carbonate peak on the HEON suggests that the concentration of basic active sites is higher in this material. These DRIFT experiments indicate the higher capability of the HEON for physisorption and chemisorption of $CO_2$ compared to the HEO and P25 $TiO_2$.

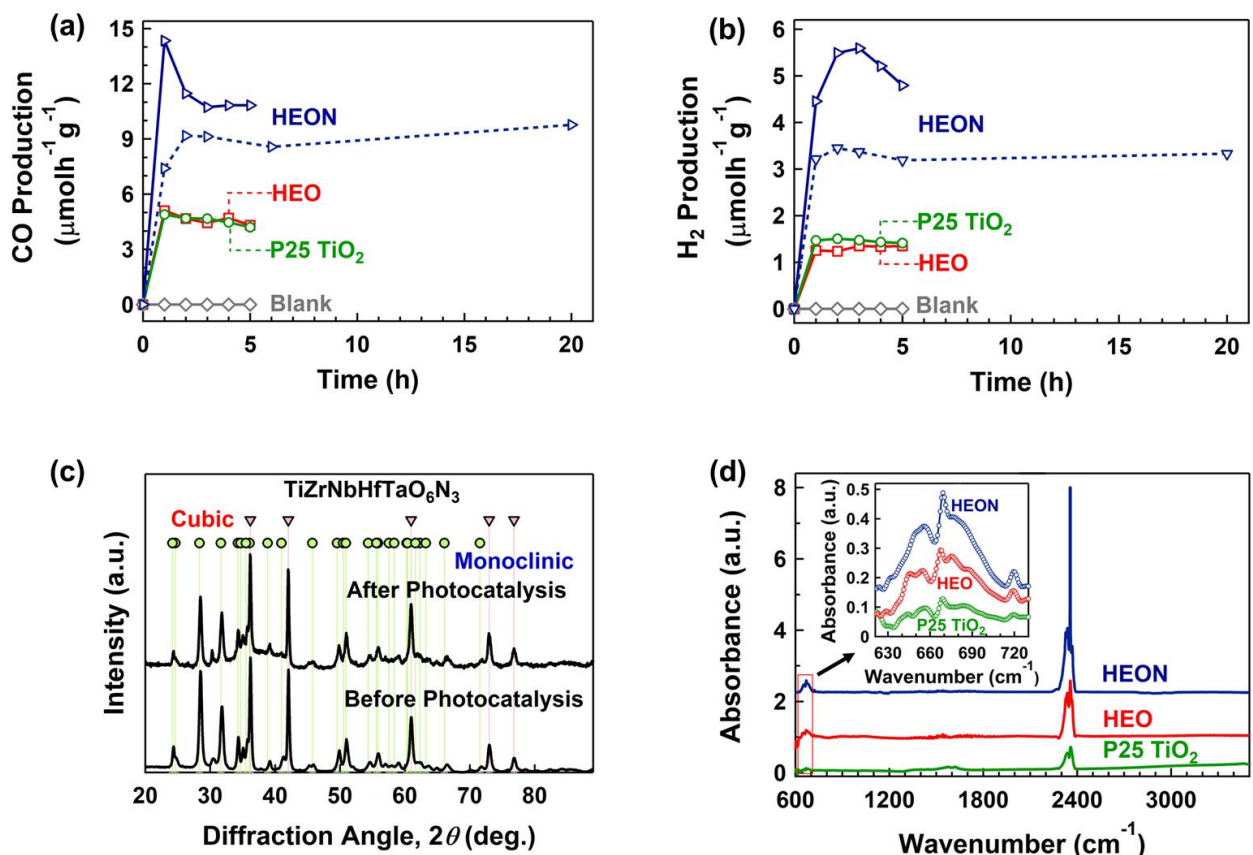

Fig. 4. High efficiency of high-entropy oxynitride for photocatalytic CO and hydrogen production. Rate of (a) $CO_2$ to CO photoreduction and (b) hydrogen generation versus UV irradiation time for high-entropy oxynitride (HEON) compared to corresponding high-entropy oxide (HEO) and P25 $TiO_2$. (c) XRD profiles before and after photocatalysis for high-entropy oxynitride. (d) DRIFT spectra of three samples.

## 4. Discussion

Three points need to be discussed in detail here: (i) the mechanism of photocatalytic CO production, (ii) the reasons for the high activity of the HEON, and (iii) the comparison of the activity of current HEON with other photocatalysts reported so far in the literature.

Regarding the first issue, it should be noted that the first step in photocatalytic $CO_2$ reduction process is the formation of $CO_2^{\bullet-}$ intermediate which is produced by sharing the electrons between $CO_2$ and photocatalyst surface [46]. Chemisorption of $CO_2$ molecules on photocatalyst surface to produce $CO_2^{\bullet-}$ occurs in three modes. (1) Nucleophilic bonding between oxygen atoms and catalyst surface (oxygen coordination), (2) electrophilic bonding between carbon atoms and catalyst surface (carbon coordination) and (3) mixed coordination between both oxygen and carbon atoms in $CO_2$ molecules with catalyst surface [46]. The chemistry of the catalyst influences the bonding of $CO_2^{\bullet-}$ with catalyst surface and determine the reaction pathway [46]. If a photocatalyst contains Sn, Pb, Hg, In, and Cd metals, then it has a tendency to oxygen coordination to produce $^{\bullet}OCHO$ as an intermediate and formic acid (HCOOH) as the final product. Photocatalysts containing noble and transition metals have a tendency to carbon coordination which leads to producing $^{\bullet}CO$ and $^{\bullet}OCHO$ as intermediates. Since the bonding between $^{\bullet}CO$ and catalyst surface is weak, CO is usually the main product in this coordination. If a photocatalyst consists of Cu atoms, then $^•$CO and $^•$OCHO are produced as intermediates, but because of the strong bonding between $^•$CO and Cu, other hydrocarbons such as methane ($CH_4$) and ethanol ($C_2H_5OH$) are usually formed as final products [46]. In this study, since the HEON, the HEO and P25 $TiO_2$ consist of transition metals, the carbon coordination pathway occurs for these photocatalysts which leads to CO production. This pathway has the following reactions [46].

$$CO_2 + e^- \rightarrow CO_2^{•-} \tag{1}$$

$$CO_2^{•-} + 2e^- + 2H^+ \rightarrow CO + H_2O \tag{2}$$

Regarding the second issue, it should be considered that combining the concepts of metal oxynitrides and high-entropy ceramics was the spark starter of this study. Metal oxynitrides have been introduced as promising low-bandgap photocatalysts particularly for water splitting [9], while their application for $CO_2$ conversion is still in the initial steps [10-12]. The reason for the low bandgap of oxynitrides compared to oxide photocatalysts is that the valence band top of these materials is generated using hybridized 2p oxygen and nitrogen orbitals, but it is generated using only 2p oxygen orbitals in oxides. Since the energy level for nitrogen 2p orbitals is higher than oxygen 2p orbitals, the bandgap of oxynitrides is smaller than oxides [9]. However, these metal oxynitrides suffer from significant recombination of charge carriers and modest stability [10,13].

High-entropy ceramics are promising new materials with interesting properties because of the presence of multiple elements which leads to superior stability, large lattice defects/strain and heterogenous valence electron distribution [14,15]. The presence of various elements in the lattice of high-entropy ceramics leads not only to high configurational entropy and resultant high chemical stability for catalysis but also to lattice distortion and formation of inherent point defects such as vacancies which can act as active sites for catalysis [47]. Although future theoretical studies are required to determine the active sites in high-entropy photocatalysts, it was shown in conventional photocatalysts that vacancies on the surface adsorb $CO_2$ and activate it by decreasing the bonding energy between carbon and oxygen [42,46]. These vacancies trap electrons and act as active sites and lead to the improved photocatalytic activity for $CO_2$ reduction, but their fraction should be optimized to achieve the highest activity and best reaction selectivity [48,49].

The combination of the two concepts of oxynitrides and high-entropy ceramics led to the introduction of $TiZrNbHfTaO_6N_3$ as a highly stable and low-bandgap photocatalyst for $CO_2$ conversion with much better photocatalytic performance compared with P25 $TiO_2$. Such a high activity is particularly interesting because the surface area of the HEON is much smaller than P25 $TiO_2$: 2.3 $m^2g^{-1}$ for the HEON and 38.7 $m^2g^{-1}$ for P25 $TiO_2$ measured by the Brunauer-Emmett-Teller (BET) technique in the nitrogen atmosphere. The high activity of the HEON for $CO_2$ photoreduction can be attributed to high light absorbance (i.e., easy electron-hole separation), appropriate band positions compared to chemical potentials for reactions, and low electron-hole recombination, and high surface $CO_2$ adsorption [1,2]. The presence of interphase boundaries in this HEON can also partly contribute to the easy separation of charge carriers and improvement of photocatalytic activity [1].

Regarding the third issue, although the comparison between the current HEON and P25 $TiO_2$ using similar experimental procedures confirms the high photocatalytic activity of the HEON, it is worth comparing the activity of this HEON with the given data in the literature. Photocatalysis in various studies is performed in different conditions in terms of photoreactor type, temperature, catalyst concentration, $CO_2$ flow rate, type of light source and concentration of reactants, and thus, a comparison between different studies should be evaluated with care. The CO production rate per catalyst mass and catalyst surface area are given in Table 1 in comparison with reported photocatalysts in the literature [49-80]. Since photocatalysis occurs on the surface, normalizing the CO production rate per surface area should be more reasonable for comparison purposes. According to Table 1, the rate of CO production in the literature fluctuates in the 0.00095-1.33 $\mu molh^{-1}m^{-1}$ range, while the CO generation rate of 4.66 ± 0.3 $\mu molh^{-1}m^{-1}$ on the HEON is higher than all these reported data. Moreover, the HEON shows much better activity per both surface area and mass unit compared to other oxynitrides reported in the literature. Although these findings introduce HEON as the most effective photocatalyst for $CO_2$ photoreduction, future studies should focus on decreasing the particle size of these materials to increase their specific surface area.

**Table 1.** Photocatalytic $CO_2$ to CO conversion rate on high-entropy oxynitride compared to reported photocatalysts. For some catalysts which surface area was not reported in literature, CO production rate in $\mu molh^{-1}m^{-1}$ was not given.

| Photocatalyst | Catalyst Concentration | Light Source | CO Production Rate ($\mu molh^{-1}g^{-1}$) | CO Production Rate ($\mu molh^{-1}m^{-1}$) | Ref. |
|---|---|---|---|---|---|
| TiO₂ / Carbon Nitride Nanosheet | 25 mg (Gas System) | 150 W Xenon | 2.04 | ---- | [50] |
| TiO₂ / Graphitic Carbon | 100 mg (Gas System) | 300 W Xenon | 10.16 | 0.04 | [51] |
| TiO₂ Nanosheets Exposed {001} Facet | 1 $gL^{-1}$ (Liquid system) | Two 18 W Low-Pressure Mercury | 0.12 | 0.00095 | [52] |
| TiO₂ / CoOx Hydrogenated | 50 mg (Gas System) | 150 W UV | 1.24 | 0.0045 | [53] |
| TiO₂ 3D Ordered Microporous / Pd | 100 mg (gas system) | 300 W Xenon | 3.9 | 0.066 | [54] |
| Pt²⁺—Pt⁰ / TiO₂ | 100 mg (Gas System) | 300 W Xenon | ~12.14 | 0.7 | [55] |
| Anatase TiO₂ Hierarchical Microspheres | 200 mg (Gas System) | 40 W Mercury UV | 18.5 | 0.37 | [56] |
| TiO₂ and Zn(II) Porphyrin Mixed Phases | 60 mg (Gas System) | 300 W Xenon | 8 | 0.062 | [57] |
| Anatase TiO₂ Hollow Sphere | 100 mg (Gas System) | 40 W Mercury UV | 14 | 0.16 | [58] |
| Anatase TiO₂ Nanofibers | 50 $gL^{-1}$ (Liquid System) | 500 W Mercury Flash | 40 | ----- | [59] |
| C₃N₄ by Thermal Condensation | 100 mg (Gas System) | 350 W Mercury | 4.83 | ------ | [60] |
| Cd₁₋ₓZnₓS | 45 mg (Gas System) | UV-LED irradiation | 2.9 | 0.015 | [61] |
| BiOI | 150 mg (Gas System) | 300 W High-Pressure xenon | 4.1 | 0.03 | [62] |
| xCu₂O / Zn₂.₂ₓCr | 4 $gL^{-1}$ (Liquid System) | 200 W Mercury-Xenon | 2.5 | 0.018 | [63] |
| CeO₂₋ₓ | 50 mg (Gas System) | 300 W Xenon | 1.65 | 0.08 | [64] |
| Cu₂O / RuOₓ | 500 mg (Gas System) | 150 W Xenon | 0.88 | --- | [65] |
| Bi₂Sn₂O₇ | 0.4 $gL^{-1}$ (Liquid System) | 300 W xenon | 14.88 | 0.24 | [66] |
| Ag / Bi / BiVO₄ | 10 mg (Gas System) | 300W Xenon Illuminator | 5.19 | 0.42 | [67] |
| g-C₃N₄ / BiOCl | 20 mg (Gas System) | 300 W High-Pressure Xenon | 4.73 | --- | [68] |
| Fe / g-C₃N₄ | 1 $gL^{-1}$ (Liquid System) | 300 W Xenon | ~22.5 | 0.06 | [69] |
| Bi₂MoO₆ | 0.7 $gL^{-1}$ (Liquid System) | 300 W Xenon | 41.5 | 1.26 | [70] |
| Bi₂₄O₃₁Cl₁₀ | 50 mg (Gas System) | 300 W High-Pressure Xenon | 0.9 | --- | [71] |
| g-C₃N₄ / Zinc Carbodiimide / Zeolitic Imidazolate Framework | 100 mg (Gas System) | 300 W Xenon | ~0.45 | 0.014 | [72] |
| BiVO₄ / C / Cu₂O | --- | 300 W Xenon | 3.01 | ---- | [73] |
| g-C₃N₄ / α-Fe₂O₃ | 200 mg (Gas System) | 300 W Xenon | 5.7 | ----- | [74] |
| Bicrystalline Anatase/Brookite TiO₂ Microspheres | 30 mg (Gas System) | 150 W Solar Simulator | 145 | 0.95 | [75] |
| 10 wt % In-Doped Anatase TiO₂ | 250 mg (Gas System) | 500 W Mercury Flash | 81 | 1.33 | [76] |
| 10 wt % Montmorillonite-Loaded TiO₂ | 50 mg (Gas System) | 500 W Mercury | 103 | 1.25 | [77] |
| Bi₄O₅Br₂ | 20 mg (Gas System) | 300 W High-Pressure Xenon | 63.13 | 0.58 | [78] |
| ZnGaON | --- | 1600 W Xenon | 1.05 | --- | [79] |
| WO₃ / LaTiO₂N | 10 mg (Gas System) | 300 W Xenon | 2.21 | 0.4 | [80] |
| α-Fe₂O₃ / LaTiO₂N | 20 mg (Gas System) | 300 W Xenon | 9.7 | 0.65 | [10] |
| RuRu / Ag / TaON | 1 $gL^{-1}$ (Liquid System) | High-Pressure Mercury | 5 | ---- | [11] |
| RuRu / TaON | 1 $gL^{-1}$ (Liquid System) | High-Pressure Mercury | 3.33 | ---- | [11] |
| Ag / TaON / RuBLRu′ | 2 $gL^{-1}$ (Liquid System) | 500 W High Pressure Mercury | 0.056 | ---- | [12] |
| TiZrHfNbTaO₆N₃ | 0.2 $gL^{-1}$ (Liquid System) | 400 W High-Pressure Mercury | 10.7 ± 1.8 | 4.7 ± 0.3 | This study |

## 4. Conclusion

This study introduced a high-entropy oxynitride with low bandgap, low electron-hole recombination, high $CO_2$ adsorbance and high chemical stability for photocatalytic $CO_2$ conversion. The material, which was synthesized using high-pressure torsion and subsequent oxidation and nitriding, had two phases of face-centered cubic and monoclinic with a chemical composition of $TiZrNbHfTaO_6N_3$. The material had better photocatalytic $CO_2$ conversion performance compared with corresponding high-entropy oxide, benchmark photocatalyst P25 $TiO_2$ and all reported photocatalysts in the literature. These findings open a new path to developing highly efficient photocatalysts for $CO_2$ conversion.

## Acknowledgments

This study is supported partly by the Mitsui Chemicals, Inc., Japan, partly by Hosokawa Powder Technology Foundation, Japan, and partly through Grants-in-Aid from the Japan Society for the Promotion of Science (JSPS), Japan (JP19H05176 & JP21H00150).